\documentclass{article}

\usepackage{multicol} 
\usepackage[svgnames]{xcolor} 

\usepackage{times} 

\usepackage{graphicx} 
\graphicspath{{figures/}} 
\usepackage{booktabs} 
\usepackage[font=small,labelfont=bf]{caption} 
\usepackage{amsfonts, amsmath, amsthm, amssymb} 
\usepackage{wrapfig} 

\title{Kaleidoscope of Quantum Coherent States}

\author{Oktay K. Pashaev and  Ayg\"{u}l Ko\c{c}ak\\ \\
Department of Mathematics \\
Izmir Institute of Technology \\
 Izmir, 35430, Turkey}

\begin{document}

\maketitle

\begin{abstract}
The Schr\"odinger cat states, constructed from Glauber coherent states and applied for description of qubits, are  generalized to the kaleidoscope of coherent states related with regular n-polygon symmetry and the roots of unity. The cases of the trinity states and the quartet states are described in details. Normalization formula for these states requires introduction of specific combinations of exponential functions with mod 3 and mod 4 symmetry.
We show that for an arbitrary $n$, these states can be generated by the Quantum Fourier transform and can provide qutrits, ququats and in general, qudit  units of quantum information. Relations with quantum groups and quantum calculus are discussed.

\end{abstract}



\let\thefootnote\relax\footnote{Extended version of poster presentation in "Quantum foundations summer school" and "Contextuality
workshop", ETH Zurich, Switzerland, 18-23 June, 2017; "Mathematical Aspects of Quantum
Information", Cargese, France, 4-8 September 2017}



\section{Introduction}
\color{Black} 
The Schr\"odinger cat states as superposition of  Glauber, optical  coherent states with opposite phases 
can be considered as qubits, a unit of quantum information.  This construction can be generalized to the kaleidoscope of coherent states related with regular n-polygon symmetry and the roots of unity. Superposition of coherent states with such symmetry provides the set of orthonotmal quantum states, 
 as a description of  qutrits, ququats and qudits. It was shown recently that such quantum states as a units of quantum information processing, have advantage in secure quantum communication. Our consideration here is motivated by ideas of symmetry and $q$-calculus. As was shown in our papers \cite{p14}, \cite{p15}, \cite{p16}, application of method of images in hydrodynamics in wedge domain requires construction of $q$-periodic functions with $q$ as a root of unity and self-similar $q$-periodic functions. This construction can be considered as a discrete Fourier transform in space of complex analytic functions. Extension of these ideas to the Hilbert space for the coherent states results in a construction which we presents below.

\subsection{Glauber Coherent States}
\color{Black} 
\textbf{\textcolor[rgb]{0.00,0.07,1.00}{Heinsenberg-Weyl Algebra}$\rightarrow $ Bosonic Algebra:} $\left[\hat{a},\hat{a}^{\dag}\right]=\widehat{\textrm{I}}$
\begin{eqnarray}
\textbf{\textrm{Annihilation operator}}&:& \, \hat{a}| 0 \rangle = 0 \,\, , \, \textrm{where} \,\, | 0 \rangle \in \cal H \\
 \textbf{\textrm{Creation operator}}&:& \, \hat{a}^{\dag}| 0 \rangle = | 1 \rangle  \Rightarrow  | n\rangle=\frac{( \hat{a}^{\dag} )^n}{\sqrt{n!}} | 0 \rangle
\end{eqnarray}

Coherent States are eigenstate of annihilation operator : $ \, \hat{a}| \alpha \rangle = \alpha | \alpha \rangle , \, \alpha \in \mathbb{C}$. This gives us relation between complex plane and Hilbert space such that $\, \alpha \in \mathbb{C} \leftrightarrow | \alpha \rangle \in \cal H  $. \newline




{\color{Navy}
\textbf{Representation of coherent states in the Fock space basis:}} $D(\alpha)=e^{ \alpha \hat{a}^{\dagger}-\bar{\alpha} \hat{a}}\Rightarrow
|\alpha \rangle=e^{ \alpha \hat{a}^{\dagger}-\bar{\alpha} \hat{a}}|0\rangle $, where $D(\alpha)$ is displacement operator $\Rightarrow$
\begin{equation}
 | \alpha \rangle =e^{ -\frac{1}{2}|\alpha|^2 } \sum _{n=0}^{\infty} \frac{\alpha^n}{\sqrt{n!}} | n \rangle
\end{equation}

Inner product of coherent states: $ \langle \alpha|\beta \rangle= e^{-\frac{1}{2}|\alpha|^2-\frac{1}{2}|\beta|^{2}+\bar{\alpha}\beta} \Rightarrow$ Coherent states are not orthogonal states. Our aim is by using coherent states to construct orthogonal set of states with discrete regular polygon symmetry.
\section{Schr\"{o}dinger's Cat States}
In description of Schr\"{o}dinger cat states one introduces two orthogonal states, which are called even and odd cat states. These cat states are superposition of $| \alpha \rangle$ $\&$ $|-\alpha \rangle$:
$$| cat_{e}\rangle \equiv |\alpha_{+} \rangle \sim | \alpha \rangle + |-\alpha\rangle  \quad  \quad |cat_{o}\rangle \equiv |\alpha_{-} \rangle   \sim | \alpha \rangle - |-\alpha\rangle$$
\vspace{0.2cm}
\hspace{0.3cm}These states can be considered as a superposition of two coherent states related by rotation to angle $\pi$, which corresponds to primitive root of unity $q^2 =\overline{q}\,^2= -1$, so that $q^4=1$ and normalization constants are $N_{+},N_{-}$:
\begin{equation}
 |0\rangle_{\textcolor{blue}{\alpha}}= N_{+}(\, |\alpha \rangle + |q^2\alpha\rangle\, )  \quad  ,  \quad |1\rangle_{\textcolor{blue}{\alpha}} = N_{-}(\,| \alpha \rangle + \overline{q}\,^2 |q^2\alpha\rangle\, )
\end{equation}
\begin{equation}
N_{+}=\frac{e^{\frac{|\alpha|^{2}}{2}}}{2\sqrt{cosh|\alpha|^{2}}} \quad \textbf{\&}\quad
N_{-}=\frac{e^{\frac{|\alpha|^{2}}{2}}}{2\sqrt{sinh|\alpha|^{2}}}
\end{equation} \\
\hspace{0.3cm}We can represent these states in matrix form by acting with  Hadamard gate:\\
\begin{equation}
\left[
    \begin{array}{c}
      |0\rangle_{\textcolor{blue}{\alpha}} \\
      |1\rangle_{\textcolor{blue}{\alpha}} \\
    \end{array}
  \right] =\textbf{\textcolor{purple}{N}}\underbrace{\frac{1}{\sqrt{2}}\left[\begin{array}{cc}
                                                1 & 1 \\
                                                  1 & \overline{q}\,^2 \\
                                                \end{array}
                                              \right]}_{Hadamard\,\,gate}\left[
                                                       \begin{array}{c}
                                                         | \alpha \rangle\\
                                                         | q^2\alpha \rangle \\
                                                       \end{array}
                                                     \right]
                                                     \end{equation}

\begin{equation}
\textcolor{purple}{\textbf{N}}=\frac{e^{\frac{|\alpha|^{2}}{2}}}{\sqrt{2}}\,\textrm{diag}\left( _{0}e^{|\alpha|^2}, _{1}e^{|\alpha|^2}\right)^{-1/2}(mod \,2)
\end{equation}
where
\begin{equation}
(mod \,2)\quad  _0 e^{|\alpha|^2}    \equiv\sum _{k=0}^{\infty} \frac{\left(|\alpha|^{2}\right)^{2k}}{(2k)!}=\frac{e^{|\alpha|^{2}}+e^{{q}^{2}|\alpha|^{2}}}{2}
\end{equation}
and
 \begin{equation}
(mod \,2) \quad _1 e^{|\alpha|^2}   \equiv\sum _{k=0}^{\infty} \frac{\left(|\alpha|^{2}\right)^{2k+1}}{(2k+1)!}=\frac{e^{|\alpha|^{2}}+ \bar q^2 e^{{q}^{2}|\alpha|^{2}}}{2}
\end{equation}

These state have been used as superposition of optical coherent states for description of qubit in quantum information processing.

\section{Trinity States}

\hspace{0.3cm}To construct three orthonormal states we consider three coherent states rotated to angle $\frac{2\pi}{3}$ (Figure 1), which corresponds to $q^6=1$. First we define $$|0\rangle_{\alpha}=N_{0}\left(|\alpha\rangle+|q^{2}\alpha\rangle+|q^{4}\alpha\rangle \right)$$
\hspace{1 cm} and  $$\, q^{6n}=1 \, \Rightarrow $$

$$(q^{2n}-1)(1+q^{2n}+q^{4n})=0 \, \Rightarrow \,$$

\begin{equation}
1+q^{2n}+q^{4n}=3\,\delta_{n \equiv 0(mod\,3)} \qquad \label{mod3}
\end{equation}

From normalization  by using (\ref{mod3})  we introduce

$$(mod \,3) \quad\displaystyle{_{0}e^{|\alpha|^2}
\equiv \sum_{k=0}^{\infty}\frac{\left(|\alpha|^2\right)^{3k}}{(3k)!}= \frac{1}{3}\left( e^{|\alpha|^{2}}+e^{{q}^{2}|\alpha|^{2}}+e^{{q}^{4}|\alpha|^{2}} \right) }$$

\begin{center}\vspace{0.25cm}
\includegraphics[width=0.43\linewidth]{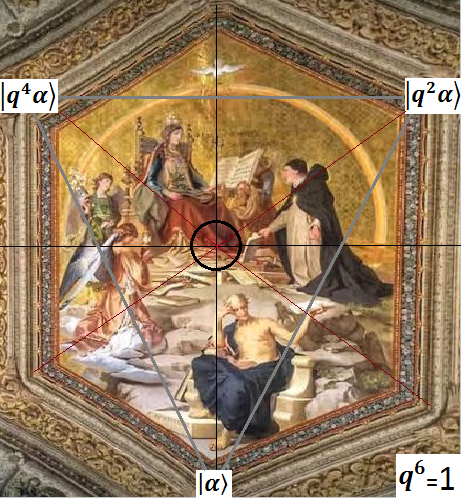}
\captionof{figure}{\color{Navy}Trinity States - ''Bene Scripsisti De Me Thoma''in Vatican}
\end{center}\vspace{0.25cm}

Then we find three orthonormal basis states as $|0\rangle_{\alpha},|1\rangle_{\alpha} \, \& \, |2\rangle_{\alpha}$:\\
\begin{eqnarray}
 |0\rangle_{\alpha}&=&e^{\frac{|\alpha|^{2}}{2}} \frac{| \alpha \rangle+|q\,^{2}\alpha\rangle+|q\,^{4}\alpha\rangle}
{\sqrt{3}\sqrt{e^{|\alpha|^{2}}+e^{{q}^{2}|\alpha|^{2}}+e^{{q}^{4}|\alpha|^{2}}}}
 =e^{\frac{|\alpha|^{2}}{2}} \frac{| \alpha \rangle + |q\,^{2}\alpha\rangle+|q\,^{4}\alpha\rangle }{3\sqrt{ _{0}e^{|\alpha|^2}{(mod \,3)}}} \nonumber \\ \nonumber \\
 |1\rangle_{\alpha}&=& e^{\frac{|\alpha|^{2}}{2}} \frac{| \alpha \rangle+\overline{q}^{2}|q\,^{2}\alpha\rangle+\overline{q}^{4}|q\,^{4}\alpha\rangle}
{\sqrt{3}\sqrt{e^{|\alpha|^{2}}+\overline{q}^{2}e^{{q}^{2}|\alpha|^{2}}+\overline{q}^{4}e^{{q}^{4}|\alpha|^{2}}}}
= e^{\frac{|\alpha|^{2}}{2}} \frac{| \alpha \rangle +\overline{q}^{2} |q\,^{2}\alpha\rangle+
 \overline{q}^{4}|q\,^{4}\alpha\rangle }{3\sqrt{ _{1}e^{|\alpha|^2}(mod \,3)}} \nonumber \\ \nonumber\\
|2\rangle_{\alpha}&=& e^{\frac{|\alpha|^{2}}{2}} \frac{| \alpha \rangle+\overline{q}^{4}|q\,^{2}\alpha\rangle+\overline{q}^{2}|q\,^{4}\alpha\rangle}
{\sqrt{3}\sqrt{e^{|\alpha|^{2}}+\overline{q}^{4}e^{{q}^{2}|\alpha|^{2}}+\overline{q}^{2}e^{{q}^{4}|\alpha|^{2}}}}
= e^{\frac{|\alpha|^{2}}{2}} \frac{| \alpha \rangle +\overline{q}^{4} |q\,^{2}\alpha\rangle+
 \overline{q}^{2}|q\,^{4}\alpha\rangle }{3\sqrt{ _{2}e^{|\alpha|^2}(mod \,3)}}\nonumber
\end{eqnarray} \\
\textbf{\textcolor[rgb]{0.00,0.07,1.00}{Matrix form of Trinity states}:} \\
\begin{equation}
\left[
  \begin{array}{c}
    |0\rangle_{\alpha} \\
    |1\rangle_{\alpha} \\
    |2\rangle_{\alpha} \\
  \end{array} \right]= \textbf{N}\underbrace{\textcolor{purple}{\frac{1}{\sqrt{3}} \left[ \begin{array}{ccc}
                                                                     1 & 1 & 1 \\
                                                                     1 & \overline{q}^{2} & \left(\overline{q}^{2}\right)^{2} \\
                                                                     1 & \overline{q}^{4} & \left(\overline{q}^{4}\right)^{2} \\
                                                                   \end{array}
                                                                 \right]}}_{Trinity\,\,gate}\left[
                                                                          \begin{array}{c}
                                                                             | \alpha \rangle\\
                                                                             | {q}^{2}\alpha \rangle \\
                                                                             | {q}^{4}\alpha \rangle\\
                                                                          \end{array}
                                                                        \right] \quad ,
\end{equation}

\begin{equation}
  \textbf{N}=\frac{e^{\frac{|\alpha|^{2}}{2}}}{\sqrt{3}}\,\textrm{diag}\left( _{0}e^{|\alpha|^2}, _{1}e^{|\alpha|^2},_{2}e^{|\alpha|^2}\right)^{-1/2}(mod\, 3)
\end{equation}

$$\boxed{  1+\overline{q}^{2(n-k)}+\overline{q}^{4(n-k)}=3\,\delta_{n\equiv k(mod \,3)}\quad , 0\leq k \leq2 }$$

These states can be used as qutrits quantum information states, having advantage in secure quantum communications.

\section{Quartet States}

\begin{center}\vspace{0.25cm}
\includegraphics[width=0.42\linewidth]{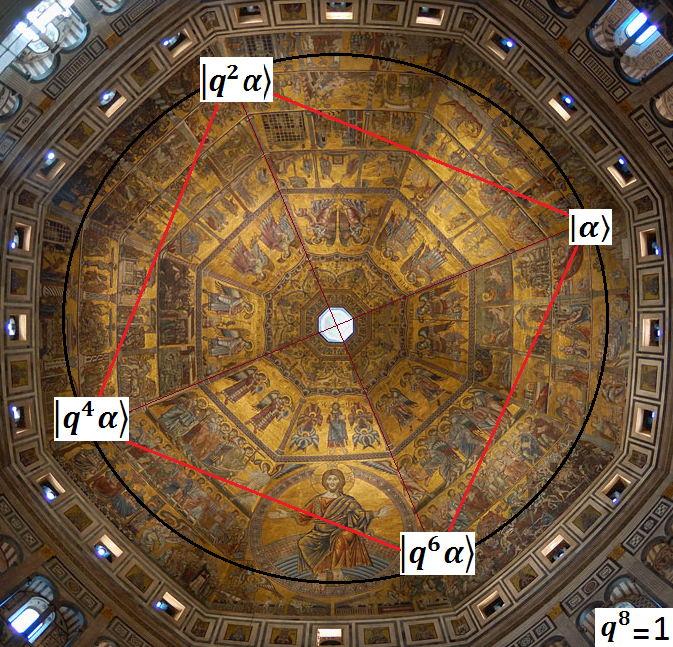}
\captionof{figure}{\color{Navy}Quartet States - Ceiling Mosaic of San Giovanni in Florence }
\end{center}\vspace{0.25cm}

\hspace{0.3cm} In Figure 2, we have four states  rotated by  angle $\frac{\pi}{2}$, determined by primitive root $$q^8 =1$$

 Superposition of these states with proper coefficients give us quartet of basis orthogonal states:
\begin{equation}
  \left[
    \begin{array}{c}
      |0\rangle_{\alpha} \\
      |1\rangle_{\alpha} \\
      |2\rangle_{\alpha} \\
      |3\rangle_{\alpha}\\
    \end{array}
  \right]=\textbf{N}\underbrace{\textcolor{purple}{\small{ \frac{1}{\sqrt{4}}\left[\begin{array}{cccc}
                                                      1 & 1 & 1 & 1 \\
                                                      1 & \overline{q}^{2} & \left(\overline{q}^{2}\right)^{2} &\left(\overline{q}^{2}\right)^{3} \\
                                                      1 & \overline{q}^{4} & \left(\overline{q}^{4}\right)^{2} & \left(\overline{q}^{4}\right)^{3} \\
                                                      1 & \overline{q}^{6}& \left(\overline{q}^{6}\right)^{2} & \left(\overline{q}^{6}\right)^{3} \\
                                                    \end{array}
                                                  \right]}}}_{Quartet\,\,gate}\left[
                                                           \begin{array}{c}
                                                             | \alpha \rangle \\
                                                             | {q}^{2}\alpha \rangle \\
                                                             | {q}^{4}\alpha \rangle \\
                                                             | {q}^{6}\alpha \rangle \\
                                                           \end{array} \right]
\end{equation}

where $$\,\, \displaystyle{ \textbf{N}=\frac{e^{\frac{|\alpha|^{2}}{2}}}{\sqrt{4}}\,\textrm{diag}\left( _{0}e^{|\alpha|^2}, _{1}e^{|\alpha|^2},_{2}e^{|\alpha|^2},_{3}e^{|\alpha|^2}\right)^{-1/2}(mod\, 4) }$$
$$\boxed{  1+\overline{q}^{2(n-k)}+\overline{q}^{4(n-k)}+\overline{q}^{6(n-k)}=4\,\delta_{n\equiv k(mod \,4)} \quad ,0\leq k \leq3 }$$

These states can be used as ququats quantum information states, having advantage in secure quantum communications.

\section{Generalized n-Cat States}

Consider superposition of $n$ coherent states, which are belonging to vertices of regular $n$-polygon and  rotated by angle $\frac{\pi}{n}$ (Figure 3).  It is related with primitive root of unity: $$q^{2n}=1$$

 $\textbf{Inner product of $q^{2k}$}\,\, \textbf{rotated coherent states:}$
 \vspace{1cm}
\begin{itemize}
  \item $\langle q^{2k}\alpha |q^{2k}\alpha \rangle =1 ,\quad   0\leq k \leq n-1$
  \item $\langle q^{2k}\alpha |q^{2l}\alpha \rangle =e^{|\alpha|^2({q}^{2(l-k)}-1)}   ,\quad  0\leq k,l \leq n-1 $
\end{itemize}

\begin{center}\vspace{0.25cm}
\includegraphics[width=0.90\linewidth]{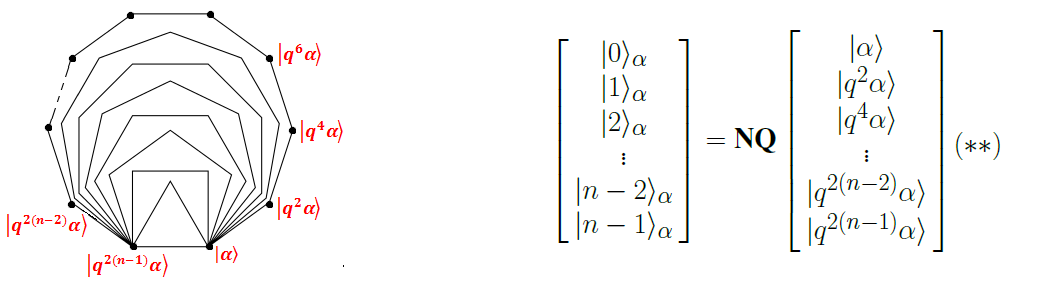}
\captionof{figure}{\color{Navy}n-regular polygon }
\end{center}\vspace{0.25cm}

\textbf{Lemma:} For $q^{2n}=1 \, , \, 0\leq s \leq n-1$
\begin{itemize}
  \item $1+q^{2m}+q^{4m}+...+q^{2m(n-1)}= n\delta_{m\equiv0(mod\, n)}$
  \item $1+q^{2(m-s)}+q^{4(m-s)}+...+q^{2(m-s)(n-1)}= n\delta_{m\equiv s(mod \,n)}$
\end{itemize}

{\color{Blue} 
\subsection{Quantum Fourier Transformation} } Construction $(**)$, shows that our orthogonal states can by described by the Quantum Fourier transform:
\begin{equation}
\left[
   \begin{array}{c}
     |\widetilde{0}\rangle_{\alpha} \\
     |\widetilde{1}\rangle_{\alpha} \\
     |\widetilde{2}\rangle_{\alpha} \\
     |\widetilde{3}\rangle_{\alpha} \\
     \vdots \\
   \tiny{ |\widetilde{n-1}\rangle_{\alpha} } \\
   \end{array}
 \right]=\frac{1}{\sqrt{n}}\left[
                    \begin{array}{cccccccc}
1 & 1 & 1 & ... & 1 \\
1 & w & w^{2} & ... & w^{n-1}  \\
1 & w^{2} & w^{4} & ... & w^{2(n-1)}\\
1 & w^{3} & w^{6} & ... & w^{3(n-1)}\\
\vdots & \vdots & \vdots & \ddots & \vdots \\
1 & w^{(n-1)}  & w^{2(n-1)} & ...  & w^{(n-1)(n-1)} \\
                             \end{array}
                           \right] \left[
   \begin{array}{c}
     |\alpha\rangle \\
     |q^2\alpha\rangle \\
     |q^4\alpha\rangle \\
     |q^6\alpha\rangle \\
     \vdots \\
    \tiny{ |q^{2(n-1)}\alpha\rangle }\\
   \end{array}
 \right]
\end{equation}
This matrix corresponds to the Quantum Fourier transform, where $w=e^\frac{-2\pi i}{\,n}$ is a n-th rooth of unity, so that it is unitary matrix satisfying $QQ^{\dag}=Q^{\dag}Q=I$:
\begin{equation}\nonumber
\boxed{|\widetilde{k}\rangle_{\alpha}\longmapsto\frac{1}{\sqrt{n}}\sum_{j=0}^{n-1}w^{jk}|q^{2j}\alpha\rangle \quad 0\leq k \leq n-1}
\end{equation}

In order to get orthonormal states, we define normalization matrix:
$$\,\, \displaystyle{ \textbf{N}=\frac{e^{\frac{|\alpha|^{2}}{2}}}{\sqrt{n}}\,\textrm{diag}\left( _{0}e^{|\alpha|^2}, _{1}e^{|\alpha|^2},_{2}e^{|\alpha|^2},...,_{n-1}e^{|\alpha|^2}\right)^{-1/2}(mod\, n) }$$
\begin{equation} \nonumber
 f_{s}(|\alpha|^2)=_{s}e^{|\alpha|^2}(mod\, n)\equiv \sum_{k=0}^{\infty}\frac{(|\alpha|^2)^{nk+s}}{(nk+s)!} \qquad\, , 0\leq s \leq n-1
\end{equation}
These functions are solution of the ordinary differential equation with proper initial values.

{\color{Blue} 
{\textbf{Differential equation}:} }$\boxed{
 \frac{\partial^n}{\partial (|\alpha|^{2})^n} f_{s}= f_{s} \quad ,\,\textrm{where} \,\,  0\leq s \leq n-1 }$

These states can be used as qudits quantum information states, having advantage in secure quantum communications.



\section{Conclusions}
Decomposition with Clock $\&$  Shift Matrices, such that $\displaystyle{\Sigma_{1}=Q\Sigma_{3}Q^\dag}$, relates our results with Quantum groups. By using cat states for qubit as a unit of quantum information we can apply our generalized n-cat states for information units as qutrits, ququats and qudits. Description of our kaleidoskope of coherent states can be realized by operator Quantum Calculus and Quantum Fourier transform.

\color{Black} 


\color{Black} 

\subsection*{Acknowledgements} This work is supported by TUBITAK grant 116F206.


\end{document}